# How Complete is the Electronic Catalogue Archive at CDS in the Extragalactic Domain?

H. ANDERNACH [1,2]

[1] Observatoire de Lyon, 9 av. Charles André, F-69561 Saint-Genis-Laval Cedex, France

[2] CDS, Obs. Astronomique de Strasbourg, 11 rue de l'Université, F-67000 Strasbourg, France

**Abstract:** From the literature of 1987 to 1993 a representative sample was prepared of 500 papers with tabular information for $\gtrsim$50 supposedly extragalactic objects each. Regardless of their scientific value, but corrected for redundancy, these papers contain 842,000 entries. For the 374 papers with >100 entries each, the CDS FTP-archive of catalogues contains tables for 21% of them, increasing from 11% for 1987 to 29% for 1993. The coverage in terms of entries is much higher, and half of all papers with >1500 entries are in the archive. Valuable and/or large data sets are identified among the missing ones. Some of these can be found in other archives, and it is argued that tabular material published in future should be preserved more systematically in a collaborative effort between authors, journal editors and data centres.

## 1. Rationale

Current computer networks allow almost instantaneous access to data obtained or compiled by other workers anywhere around the world (see [5] for a review). The Strasbourg Astronomical Data Centre (CDS) offers the largest collection world-wide of commonly used astronomical compilations and published tables in a publicly accessible FTP-archive, which in May 1994 comprised ~900 data sets [13]. The CDS, originally a *stellar* data centre, was renamed to "Astronomical" Data Centre in late 1991, and it now covers extragalactic data equally. Both Galactic and extragalactic objects may be extended, and published data may be classified into global (integrated), one-dimensional (spectral), two-dimensional (imaging) or even three-dimensional (imaging spectroscopy) data. The *global* data are technically the simplest to archive. They are usually the only to enter databases like SIMBAD, NED, or LEDA, and thus form their basic "food". However, such tables also allow other users to explore the data in different ways. Both these aspects motivated a quantitative study of the coverage achieved in the CDS catalogue archive (i.e. *not* that of SIMBAD) for extragalactic objects.

The test sample of references is presented in section 2, and section 3 assesses the frequency with which electronic information from these references are found at the CDS. In section 4 some other sources of electronic catalogues are described. Section 5 identifies examples of data sets missing at CDS and examines who are the customers of the CDS archive. In section 6 some conclusions are drawn and suggestions for improvements are made.

## 2.    The Test Sample of References

The aim was to assess the completeness with which major compilations, catalogues, atlases, and tables with published observational data are found in the electronic CDS archive. A minimum of ∼50 supposed extragalactic objects per paper was required for the current test sample, regardless of their *scientific value*. This number of objects was allowed to be lower if many lines (*entries*) of data were given per object. Papers dominated by one- or higher-dimensional data were only included if they also contained tabular information on global data. Thus, *e.g.* Tully's *Nearby Galaxies Atlas* was excluded, while the accompanying *Nearby Galaxies Catalogue* was included. Individual samples of ≲100 objects may be of limited scientific use, but their existence in electronic form would make the work of database managers and compilers of catalogues more efficient, *e.g.* to attach a reference to hundreds of objects for which this paper displays a spectrum. However, as the emphasis was on a *representative* sample of the literature the above conditions were not always taken as ironclad. Statistical statements will be based on a subsample of papers with over one hundred entries only.

Radio-source catalogues were also included, as the large majority of unresolved sources is *extragalactic*, even near the Galactic plane [9]. Papers dedicated to Galactic sources (planetary nebulae, HII regions or pulsars) were excluded. Optical counterparts of radio sources are frequently below the magnitude limits of current galaxy catalogues, thus only a small fraction of them (probably ≲ 15%) presently have an optical identification.

The period from 1987 to 1993 was chosen. This is long enough to study trends with time, and virtually all tabular data published since 1987 must have existed in electronic form at some stage. Instead of visually scanning the *entire* literature for such a period, I used the bibliography of atlases and catalogues as given in the IAU triennial reports [1, 11, 15] and in section 002 of *Astronomy & Astrophysics Abstracts* (AAA) up to vol. 57. Many of the IAU and AAA references (with ≲50 objects or data difficult to prepare for CDS's electronic archive) were rejected. Some journals (like A&A, ApJ and MNRAS) were clearly underrepresented in the IAU lists as compared to others (like A&AS, ApJS and AJ). Eventually I browsed several journals (especially ApJS, A&AS, AJ and MNRAS) and also filled in the second half of 1993 not covered by IAU or AAA bibliographies. Finally, I cross-checked with my own unpublished compilation of references, as well as against a list of several hundred radio-continuum surveys published before 1991 [19].

The final sample of 500 references is available on request from the author. Standard reference codes are given in the Appendix. The sample is not *complete* in the sense that the A&A and ApJ main journals were not scanned in full. Obvious redundancy among the papers was removed, *e.g.* for the quasar catalogues by Hewitt & Burbidge and Veron-Cetty & Veron only the most recent version was used. Tables from some papers may have been integrated in other compilations (*e.g.* redshift compilations like ZCAT or SRC, available as CDS catalogues VII/164 and VII/142), but this may not be an argument against having the data *as published* in the CDS catalogue collection : other users might want to use the data in a way different from the compilation catalogue, or use other parts of the data which did not even enter the compilation catalogue. I made sure that all relevant entries in the CDS archive were present in the test sample, *i.e.* any further addition of references would only *lower* the completeness levels of the CDS archive derived below.

The sample is believed to be at least 80 % complete and rich enough to allow some quantitative analysis. Table 1 gives a statistical overview. The origin of the references is given for each year in the first seven rows and four columns. The next two rows give the sum over the period 1987–93 for all references, and the last rows only for those papers with $N_{entr} > 100$ entries. The IAU and AAA references provided almost half of all references gathered in this exercise. Less than 5 % of the references are monographs or appeared in non-periodical publications.

## 3. Results

The public file `cdsarc.u-strasbg.fr:/pub/cats/cats.all`, with the contents of the CDS archive was searched for all 500 papers. As of May 21, 1994 the file listed ∼680 catalogues in sections I–VIII and ∼220 items in the "Journal-"(J–)directories [13]. Note however, that almost half of the items listed in this file for sections I–VII are not yet stored in the FTP archive because of lack of electronic documentation. These are mostly older and sometimes superseded data sets which will be incorporated upon request. They were thus considered as "available" in the present study.

Table 1: Test References and Presence in the CDS Electronic Catalogue Archive

| Year of Publication | from IAU | from AAA | from others | $N_{entr} > 50$ at CDS | $N_{entr} > 100$ at CDS |
|---|---|---|---|---|---|
| 1987 | 25 | 9 | 20 | 4/54 = 7 % | 4/38 = 11 % |
| 1988 | 14 | 4 | 28 | 5/56 = 9 % | 4/41 = 10 % |
| 1989 | 21 | 4 | 47 | 9/72 = 13 % | 9/56 = 16 % |
| 1990 | 26 | 10 | 40 | 13/76 = 4 % | 13/64 = 20 % |
| 1991 | 36 | 5 | 43 | 22/84 = 26 % | 21/60 = 35 % |
| 1992 | 29 | 6 | 34 | 13/69 = 19 % | 11/56 = 20 % |
| 1993 | 21 | 20 | 48 | 18/89 = 20 % | 18/69 = 29 % |
| 1987–93 | 172 = 34 % | 58 = 12 % | 270 = 54 % | 84/500 = 17 % | 79/374 = 21 % |

### 3.1 STATISTICS BY NUMBER OF PUBLICATIONS

The search result for *all* references and for the subset of references with $N_{entr} > 100$ is shown in the last two columns of Table 1, respectively. The fraction of papers for which data are archived at CDS increases with year of publication, for the sample with $N_{entr} > 100$ from 11 % for 1987 to 29 % for 1993, with an average of 21 % over the seven years. Larger data sets are evidently better represented (cf. also Figure 2 below). The 84 extragalactic catalogues in the CDS archive include 24 which were collected and forwarded to CDS by the present author.

### 3.2 STATISTICS BY SIZE AND JOURNAL OF THE PUBLICATIONS

For a more quantitative assessment I also recorded the number of objects contained in each of the 500 papers. Sometimes this number could only be estimated. For some papers with few objects, but many data rows (*entries*), the total number of entries was used as the relevant measure of size. Generally the references deal with extragalactic objects only, but for the two largest data sets, namely the *IRAS Faint Source Catalog* (including the *Associations*) and the *Catalog of IR Observations*, a correction for their Galactic "contamination" was made. According to an estimate by M. Schmitz (*priv. comm.*) a total of 120,000 and 70,000 entries was assumed for the latter catalogues, respectively. For the radio source catalogues this correction would have been much smaller ($\lesssim 10\,\%$) and was ignored. Again, the statistics derived is meant to be *representative* rather than *exact*. The total extragalactic content of the 500 references was found to be $\sim$842,000 entries (not necessarily different objects). The average is $\sim$1700 entries per paper, the median $\sim$200 (see also Fig. 2).

The top seven rows of Table 2 show how *all* 500 test references are distributed among the major journals for individual years (but note that A&A and ApJ were *not* systematically scanned here). The next two rows give the total number of papers $N_{ref}^{tot}$ and the total number of entries $N_{entr}^{tot}$ they contain. While $N_{ref}^{tot}$ is highest for A&AS, followed by ApJS, AJ and MNRAS, $N_{entr}^{tot}$ is highest for ApJS, followed by A&AS, MNRAS and AJ. The next two rows give the number of references $N_{ref}$ covered by the CDS archive, and the number of entries $N_{entr}$ contained in these data sets. Both $N_{ref}$(CDS) and $N_{entr}$(CDS) are higher for ApJS than for A&AS. The results of this table practically do not change if the references are restricted to those with $N_{entr} > 100$, as can be seen from Figure 2 below.

Table 2: Partition of Test References by Journal and Size

| Year | A&AS | ApJS | AJ | MNRAS | ApJ | A&A | others | Total |
|---|---|---|---|---|---|---|---|---|
| 1987 | 17 | 15 | 6 | 5 | 4 | 0 | 7 | 54 |
| 1988 | 16 | 7 | 9 | 11 | 4 | 1 | 8 | 56 |
| 1989 | 13 | 14 | 20 | 9 | 4 | 0 | 12 | 72 |
| 1990 | 16 | 17 | 10 | 11 | 4 | 1 | 17 | 76 |
| 1991 | 18 | 18 | 15 | 13 | 7 | 2 | 10 | 83 |
| 1992 | 17 | 22 | 10 | 5 | 5 | 3 | 7 | 69 |
| 1993 | 20 | 13 | 16 | 15 | 9 | 2 | 15 | 90 |
| $N_{ref}^{tot}$ | 117 | 106 | 86 | 69 | 37 | 9 | 76 | 500 |
| $N_{entr}^{tot}$ | 155,648 | 215,696 | 27,461 | 69,954 | 6,395 | 1,392 | 375,510 | 842,000 |
| $N_{ref}$ (CDS) | 16 | 21 | 8 | 15 | 3 | 3 | 18 | 84 |
| $N_{entr}$(CDS) | 106,943 | 166,533 | 6,120 | 37,327 | 787 | 473 | 306,569 | 624,752 |

The Editors of *Astronomy & Astrophysics* (A&A) and its *Supplement Series* (A&AS) agreed that "extensive tabular data", accepted for publication in A&A or A&AS after 1992, and selected either by the Editors or the authors, would be stored at CDS (see A&AS 280, E1). To assess the realization of this agreement, I extended the literature search for A&AS up to vol. 105 (1994, see last 12 entries of Appendix). As for A&AS alone, for the period when the agreement was implemented, the archive contained data from 11 of 19 references, with 42 % (3547/8485) of the entries. Missing references which in my opinion merit to be stored at CDS, are preceded with a "-"-sign in the Appendix.

Figure 1 shows the size distribution of the 512 publications (*i.e.* including the 12 from A&AS 1994). The cumulative number of papers with $> N_{entr}$ entries is plotted *vs.* $N_{entr}$. For $N_{entr} \gtrsim 100$ the curve is almost a power-law of index near $-0.7$, a manifestation of *Zipf's law* in bibliometrics [12]. The turnover of the curve for $N_{entr} \lesssim 100$ is most likely due to incomplete sampling of smaller data sets in the literature survey.

Figure 1 (see text)                    Figure 2 (see text)

Figure 2 displays the completeness of the CDS archive as function of the number of entries per paper. The solid line shows, for a given minimum $N_{entr}$, the fraction of *papers* for which electronic tables are archived at CDS. All references with >10,000 entries are available at CDS, and half of those with $N_{entr} \gtrsim 1500$ entries. The dashed curve shows, for a given minimum number of entries $N_{entr}$ per publication, the fraction of *entries* stored at CDS. The fact that the largest catalogues (*i.e.* those typically available at CDS) dominate by far the number of data entries, implies that the coverage of the CDS archive never drops below $\sim 74\%$ even when all papers down to 50 objects are considered.

### 3.3 A Byproduct: Completeness of the SIMBAD Bibliography

As a *byproduct* (but not as an assessment of SIMBAD's extragalactic data content !) the bibliographical tool `simref` was queried for those test references published in journals covered by the SIMBAD team (see SIMBAD User's Guide & Reference Manual III, p.66). Thirty references were not resolved by `simref`, and these are marked with an asterisk in the Appendix. According to [10] it is maybe not surprising that half of the missing references are of Russian or Chinese origin (both original *and* English translation were queried). The other 15 papers, however, appeared in major journals, and seven of them deal with radio sources *only*. All thirty references contain original data.

While the high level of SIMBAD's literature coverage is reassuring, a recognition of a paper by `simref` does not imply that the *data* from these papers are also offered by SIMBAD. It usually means that the reference is attached to at least those objects having a cross-identification in SIMBAD. The on-line notes appended to some of the SIMBAD bibliographic references, especially the ones with *many* objects (like *e.g.* "to be scanned" or "to be requested in electronic form"), document the ongoing efforts of the SIMBAD team to gradually recover missing electronic information, and to establish cross-identifications for these objects in SIMBAD.

Also, there is a frequent misunderstanding that all entries in the CDS catalogues would automatically be accessible through SIMBAD. This interesting feature is now under development within the project ALADIN [17], which will eventually allow an interactive overlay of entries from both SIMBAD *and* the CDS catalogues on top of the digital images of the optical sky surveys. Both SIMBAD and ALADIN would clearly benefit from a more complete coverage of published tabular data in the electronic archive of CDS.

## 4.    Other Collections and Sources of Electronic Data

Several on-line systems like *e.g.* DIRA2, ESIS, HEASARC, STARCAT, or STARLINK also offer astronomical catalogues, among which one can occasionally find valuable items not archived at CDS. Below I describe three other possible sources of catalogues.

### 4.1  The AAS CD-ROM Series

Early in 1994 the American Astronomical Society (AAS) issued the first CD-ROM of a series with bulky data published in the AAS journals ApJ, ApJS, and AJ (see ApJ 402, 1). This CD-ROM covers the year 1993, for which my test sample lists 38 references published in ApJ, ApJS and AJ. However, the CD-ROM contains data for only six (16%) of them. One of the reasons for this small coverage could be the fact that data contribution by the authors is *voluntary*. Also, one CD-ROM per year seems too infrequent. It not only burdens the authors with data requests (cf. sect. 5), it may even *inhibit* full comprehension of papers for which data are given in electronic form *only* (*e.g.* 1994AJ.107.1629T, not in the present test sample) until the CD-ROM appears. Database managers would like to use the data within weeks from publication. The amount of data on the AAS CD-ROM is small enough to be integrated in the CDS archive.

### 4.2  Radio Sources in EOLS and ADS

Within the framework of the Working Group "Radioastronomical Databases" of IAU Commission 40 [2] the present author has now collected 120 electronic radio-source catalogues, virtually all of them were not previously archived at data centres. Currently 55 of these tables (comprising 511,000 records) are searchable in the *Einstein On-line Service* EOLS [4] and NASA's *Astrophysics Data System* ADS. Separate ASCII tables are not provided by EOLS or ADS. However, their integration into the CDS archive only require a translation of the EOLS documentation files from ADS to CDS standards [14].

Table 3 compares the total number of existing electronic radio-source catalogues with those integrated in either CDS or EOLS/ADS. Note that the statistics in this table is *not* restricted to (although it is dominated by) the period from 1987 to 1993. As already seen in Figure 2 above, the archiving statistics is clearly better when the number of entries rather than the number of references is counted.

Table 3: Electronic Radio-source Catalogues in EOLS and in CDS

|                    | $N_{ref}$ | fraction | $N_{entr}$ | fraction |
| ------------------ | --------- | -------- | ---------- | -------- |
| total in existence | 120       | 100%     | 555,900    | 100%     |
| of which in EOLS   | 55        | 46%      | 510,848    | 92%      |
| of which at CDS    | 21        | 18%      | 230,600    | 41%      |

According to one of the most productive surveyors of the radio sky [7] "most extra-galactic radio astronomers live in a distant part of the universe, isolated from the rest of astronomy." His remedy to the problem, namely to make a deeper large-scale radio survey, in order to detect objects which the "normal" astronomer can study, is now being realized, and will contribute another 2 million radio sources before the year 2000 [6, 8].

### 4.3   Extragalactic Databases

Two public on-line databases exist for extragalactic objects. The *NASA/IPAC Extragalactic Database* (NED) [20] covers *all* extragalactic objects, while the *Lyon-Meudon Extragalactic Database* (LEDA) [18], covers nearby ($z \lesssim 0.2$) galaxies only and offers up to 66 different parameters per galaxy. These databases provide very valuable cross-identifications between individual catalogues, but they do not provide data tables as published, nor do they generally make use of *all* data columns, nor can the input data sets be reconstructed by database interrogation. However, both need these tables to feed their databases, and frequently these are page-scanned and must then be proof-read. Much time could be saved if database teams could obtain tables routinely from a central archive. Vice versa, major tables which NED or LEDA received from the authors or converted from paper into electronic form could be provided to that archive. The well-documented and easily searchable FTP-facility for electronic catalogues at CDS would be an obvious repository.

## 5.   Examples of "Missing" Data and their Customers

It is often argued that modern (*i.e.* larger and more sensitive) data sets would supersede previous data anyway. This is invalid if these older data were taken with different equipment, in another waveband or with different angular resolution, in addition to the different epoch, which would allow to check for variability. A complete recovery of data sets "missing" in the CDS electronic archive is equally impossible and unnecessary, but selection criteria for those to be recovered should be defined. Among the eighty references with >1000 entries I found, *e.g.* catalogues of *HI Observations of 10,000 galaxies* (AAA.49.002.026H); *6445 Southern Peculiar Galaxies* (AAA.43.002.088A); $\sim$*4000 Nearby Galaxies* (AAA.45.002.054T); 2300 *Southern Ringed Galaxies* (AAA.53.002.023B); *Visual and IR photometry of $\sim$2000 galaxies* (AAA.50.002.113D); lists of *Low Surface Brightness Dwarfs* (AAA.49.002.069K) and of a *Drift Scan Survey* (1993AJ.105.393K) with $\sim$1500 galaxies each, not to mention over a dozen radio-source catalogues. Two authors whose data I requested recently stated that they had been asked for the same data sets (1993AJ.106.1273 and 1980ApJS.42.565) over 50 resp. over 100 times (!). This proves the benefit of such a storage at CDS. Clearly, a *regular* archiving of all (even medium-sized) data sets would be essential in saving manpower needed to keep extragalactic databases complete and up-to-date, and to help compilers of reference catalogues, like *e.g.* this one:

*During my work ... I repeatedly had to refer to the observational data which involved looking through a large number of individual papers each time. ...much time would be saved if observational data were kept on file for individual sources. This Handbook is the result of the integration of a part of such a file into a volume. I hope the use of this volume will now save time for others* [16].

This work of 1978 collected for the first time the most relevant data for almost 200 strong radio sources from $0^h$ to $12^h$ of RA(1950). However, the other half of the sky, and another announced volume on weaker sources, was never finished for the obvious reason of an overwhelming effort required to compile them.

With present-day computers and networking capabilities we are clearly in a much better starting position. However, along the way of developing such tools, and even now, much information which once existed in "machine-readable" form (starting with punched cards !) was not saved on appropriate media after publication. Some of this material is now being recovered (at CDS and elsewhere) with optical character recognition (OCR) devices, but at the expense of large amounts of manpower for proofreading. Several large compilations were published in extremely small and sometimes broken fonts (many of them on microfiche), impossible to recover even with modern OCR methods. It is surprising that even for papers published during the past 9 months (in A&AS by authors in central Europe) it was necessary to page-scan the tables at CDS because the authors either did not provide the data or had lost them. A supply of the data directly to the data centre would be a minimal effort and make such exercises unnecessary.

To get an idea on who are the "customers" of CDS data, I inspected the files that record the downloading of data sets from the FTP archive for sections VII–VIII of the archive (*i.e.* for non-stellar and radio data). After exclusion of file accesses for local backups, I found that the great majority of FTP-copies were made by staff of other data centres, database managers, a few compilers of reference catalogues, but very few by other individual astronomers. This differs from the other sections of the archive: among the 27 most copied catalogues since 1991, only two are extragalactic (Ochsenbein, priv. comm.).

## 6.  Conclusions

The catalogue archive at CDS concentrates the world's largest set of astronomical catalogues in one place, and this paper provides the first quantitative assessment of its *extragalactic* content. While a clear trend for better coverage of more *recent* publications and *larger* data sets is evident, data from only 21 % of the 374 papers with $\gtrsim 100$ extragalactic entries and published between 1987 and 1993 are in the archive. This does not imply that all other material must be restored, but rather gives in impression on how the astronomical community has preserved material that was originally quite costly to prepare. For the period since implementation of the A&A–CDS agreement (see sect. 3.2), the CDS archive was found to contain 11 of 19 (extragalactic) papers with 42 % of the data entries. This is less than expected from (at least my own) judging of the scientific value and/or size of the data. Several data sets could be integrated from other electronic archives, but $\sim 140$ of the 500 references were not found in any of these. The existence of *different* sets of astronomical catalogues in different archives prompted me to suggest in 1993 a *Master Index of Astronomical Catalogues* [3] in which the CDS could play an important role.

A possible reason for the low number of FTP-copies of extragalactic catalogues might be that CDS is not yet sufficiently known as an archive of such data. Also, most astronomers are probably looking for a database which would automatically provide them with entries from *many* individual catalogues, and are much less ready to select, copy, and browse these catalogues by themselves.

Clearly, the CDS will have to cope with ever *more* data sets. Hundreds of users worldwide now rely on public databases for their bibliographical object studies. To accomplish a reasonable completeness, database managers must rely on the input of electronic tables "*as published*, since an enormous amount of time will be saved from optically scanning the tables and proofreading" [20]. More coherent collaboration is needed between authors, editors, and data centres to arrive at a satisfactory level of completeness. Above all, the authors should be more willing to supply their data directly to the data centers. Editors of journals can help "stimulate" this attitude, and editors *other* than those of A&A and A&AS should join the commitment to supply tables contained in publications to the data centres. In a suitable collaboration between data centres, these should care-fully monitor the literature for items missing in their archives and actively request these data. Eventually the *user* should be encouraged to provide quality control even years after the information was archived. A considerable boost of motivation for authors to provide their data to the CDS archive could come from the integration of more *data* (rather than only *references* attached to objects) into SIMBAD, NED or LEDA, even more so if data become transparent *graphically*, as is indeed planned with the powerful ALADIN tool.

## Acknowledgements

I am grateful to M. Crézé for suggesting this study, and for the hospitality I received at CDS where part of this work was done. F. Ochsenbein patiently explained and installed several tools facilitating this study. C. Petit page-scanned ref. [11] and S. Okamura provided ref. [15] in electronic form. The ADS abstract service saved me several visits to the library. E. Pecontal helped with the design of the figures, and valuable comments improving the manuscript were received from C. Stern Grant, M. Crézé, M. Pakull, G. Paturel, M. Schmitz, M. Kurtz, F. Simien, R.E.M. Griffin, and an anonymous referee.

**Appendix:** 500 test references 1987–1993, plus 12 papers from A&AS (1994)

See SIMBAD User's Guide III, p. 144 for the 19-digit standard reference coding. Items are sorted by year of publication, then journal. For AAA the publication year is that of the *reference*, and *not* that of the AAA volume. Other codes are: AApTr = Astronomical and Astrophysical Transactions; UCapT = Astron. Dept. Univ. Cape Town Publ.. The code is *preceded* by a "+" if stored at CDS, a \* if not recognized by the simref tool, and a "−" if published in A&AS since mid-1993 with data desirable for the CDS archive. The code is *followed* by a "−" if the paper contains <100 entries.

| | | | |
|---|---|---|---|
| 1987A&AS...67..237R | 1988A&AS...73..265A | 1989A&AS...79..283W | 1989PASP..101..360S |
| 1987A&AS...67..261R- | 1988A&AS...73..471S- | 1989A&AS...80..215A | 1989PBeiO..12....8Z |
| 1987A&AS...67..341B | 1988A&AS...73..515B | +1989A&AS...80..299P | 1990A&A...231..327R- |
| 1987A&AS...68..427R | +1988A&AS...74...83S | 1989A&AS...81..253D | 1990A&AS...82...41K |
| 1987A&AS...69...23T- | 1988A&AS...74..315R | 1989A&AS...81..291S- | 1990A&AS...82..113J |
| *1987A&AS...69...91S | 1988A&AS...74..475P | 1989AAA.49.002.026H | 1990A&AS...82..279R- |
| 1987A&AS...69..487W- | 1988A&AS...75...67L | 1989AAA.49.002.027B | +1990A&AS...82..391B |
| 1987A&AS...70...77F | 1988A&AS...75..317S | 1989AAA.49.002.128M | 1990A&AS...83..183M- |
| 1987A&AS...70...95d | 1988A&AS...76...21Q | 1989AAA.51.002.070B | 1990A&AS...83..393Q |
| 1987A&AS...70..115S- | +1988A&AS...76...65B | 1989AJ.....97...69B- | 1990A&AS...83..399G |
| *1987A&AS...70..189S | 1988A&AS...76..339M | 1989AJ.....97..315D | 1990A&AS...83..539R |
| *1987A&AS...70..191S | 1988AAA.45.002.054T | 1989AJ.....97..633G | 1990A&AS...83..569B |
| 1987A&AS...70..465d- | 1988AAA.49.002.069K | 1989AJ.....97..708v- | 1990A&AS...84...47T |
| 1987A&AS...70..517H | 1988AAA.50.002.113d | 1989AJ.....97..957H | 1990A&AS...84..455P |
| 1987A&AS...71...25O | 1988AJ.....95..284D | 1989AJ.....97.1319C | 1990A&AS...85..805F |
| *1987A&AS...71..221O | 1988AJ.....95.1340D | 1989AJ.....97.1556v | 1990A&AS...85.1049S- |
| *1987A&AS...71..493K | 1988AJ.....95.1659M | 1989AJ.....97.1576M | 1990A&AS...86..109G |
| +1987AAA.43.002.088C | 1988AJ.....95.1678M | 1989AJ.....97.1721C | 1990A&AS...86..167J |
| 1987AAA.43.002.137A | 1988AJ.....96...30C | 1989AJ.....98...54J | 1990A&AS...86..473F |
| 1987AAA.45.002.053L | +1988AJ.....96..816C | 1989AJ.....98...64Z | +1990AAA.50.002.111L |
| 1987AAA.46.002.153S | 1988AJ.....96.1655G | 1989AJ.....98..351M | 1990AAA.52.002.009K |
| 1987AJ.....93....1B- | 1988AJ.....96.1775O | +1989AJ.....98..367F | 1990AAA.52.002.059B |
| 1987AJ.....93..788D | 1988AJ.....96.1791F | 1989AJ.....98..419V | *1990AISA0..32...31A |
| 1987AJ.....94..111L- | *1988Afz...29..247S | 1989AJ.....98..766S | 1990AISA0..32...73A |
| 1987AJ.....94..587B- | +1988Afz...29..548L- | 1989AJ.....98..931V | 1990AJ.....99..463D |
| 1987AJ.....94.1116C | 1988Ap&SS.141..303B | 1989AJ.....98.1148M- | 1990AJ.....99.1071C |
| 1987AJ.....94.1423F | 1988ApJ...325..610H | 1989AJ.....98.1175M | 1990AJ.....99.1381V- |
| 1987AcA....37..163R | 1988ApJ...328..114P- | 1989AJ.....98.1195K | 1990AJ.....99.1435K |
| 1987ApJ...314..129C | 1988ApJ...328..530E- | 1989AJ.....98.1959F | 1990AJ.....99.1722K |
| 1987ApJ...317..102S | 1988ApJ...329..174T- | 1989AN....310....7B | 1990AJ.....99.1740H |
| 1987ApJ...320..238S | 1988ApJS...66....1U | 1989AZh....66..897P- | +1990AJ....100....1F |
| 1987ApJ...321...94D- | 1988ApJS...66..261K- | *1989Afz...31...63S | +1990AJ....100...47C |
| 1987ApJS...63..247H | 1988ApJS...66..297P | 1989ApJ...339...12H | 1990AJ....100.1028S |
| 1987ApJS...63..265W | 1988ApJS...67....1T | 1989ApJ...343..659U- | 1990AJ....100.1405W- |
| 1987ApJS...63..311H | +1988ApJS...68...91R- | 1989ApJ...345...59C | 1990AN....311....5B |
| 1987ApJS...63..515L | 1988ApJS...68..151H | 1989ApJ...347..127F- | *1990AZh....67....1A |
| +1987ApJS...63..543S | 1988ApJS...68..715K | 1989ApJS...69...65H | +1990Afz...32...29A |
| 1987ApJS...63..555S | 1988MNRAS.230....1B | 1989ApJS...69..365F- | *1990Afz...32..441S- |
| 1987ApJS...63..771G- | 1988MNRAS.230..639M- | +1989ApJS...69..763F | *1990Afz...33...89S- |
| 1987ApJS...63..803D- | 1988MNRAS.231..479S | 1989ApJS...69..809M- | 1990Afz...33..213A |
| 1987ApJS...63..809S | 1988MNRAS.231..977L | +1989ApJS...70....1A | *1990Afz...33..345A |
| 1987ApJS...64..411S- | 1988MNRAS.231.1065M- | 1989ApJS...70..271B- | 1990Afz...33..351S- |
| 1987ApJS...64..417S- | 1988MNRAS.232..111M | +1989ApJS...70..329K | 1990ApJ...354..124I |
| 1987ApJS...64..581D | 1988MNRAS.232..381M- | 1989ApJS...70..447S | 1990ApJ...357..388L |
| 1987ApJS...64..601B | +1988MNRAS.234..919H | 1989ApJS...70..479S | 1990ApJ...359....4v |
| 1987ApJS...65..485C | 1988MNRAS.234.1051G- | 1989ApJS...70..687H | 1990ApJ...365...66H |
| 1987ApJS...65..543C- | 1988MNRAS.235.1227C- | 1989ApJS...70..699Y | 1990ApJS...72..231M- |
| +1987BICDS..32...81K | 1988MNRAS.235.1313S | 1989ApJS...70..723M- | 1990ApJS...72..245S |
| 1987MNRAS.224..895W- | 1988PASP..100..452T | 1989ApJS...71..433P- | 1990ApJS...72..291S |
| +1987MNRAS.227..563L | 1988RMxAA..16..123C | 1989ApJS...71..701H- | 1990ApJS...72..433H |
| *1987MNRAS.227..607W | 1989Zelenchuk-7cm.A | 1989AuJPh..42..633S | +1990ApJS...72..471D |
| 1987MNRAS.227..705W- | +1989IRAS-GalQSOAssL | +1989MNRAS.236..171D- | +1990ApJS...72..567G |
| 1987MNRAS.229..589P | 1989A&As...77...31V | 1989MNRAS.236..207M | 1990ApJS...72..621L |
| 1987PASJ...39..709H | 1989A&As...77...75F | 1989MNRAS.236..425U | 1990ApJS...72..761B |
| 1988A&A...189...7I- | 1989A&As...77..161C | 1989MNRAS.236..737E- | 1990ApJS...73..359C |
| 1988A&AS...72..215P | 1989A&As...77..237R | 1989MNRAS.238.1171D | 1990ApJS...73..603T |
| 1988A&AS...72..243R- | 1989A&As...78..269F | 1989MNRAS.239..459D | 1990ApJS...74....1Z |
| 1988A&AS...72..415P | 1989A&As...78..277G | 1989MNRAS.240..591S | 1990ApJS...74..129G |
| 1988A&AS...73..103O | +1989A&As...79...79S | 1989MNRAS.240..657S- | 1990ApJS...74..181Z |
| 1988A&AS...73..125O- | +1989A&As...79..105S | +1989MNRAS.240..785M | 1990ApJS...74..325B |
| 1990ApJS...74..347K- | 1991ApJS...75..273O | 1992ApJ...386..143C | 1993AJ....105.1637H |
| 1990ApJS...74..675B | +1991ApJS...75..297H | 1992ApJ...386..408H- | 1993AJ....105.2107M- |
| *1990ApJS...74..869B | +1991ApJS...75..645K- | 1992ApJ...387..591J | 1993AJ....106..473C |
| *1990ChA&A..15..131H | 1991ApJS...75..751R | 1992ApJ...400..410B | 1993AJ....106..831H |
| +1990IRASF.C...0000M | 1991ApJS...75..801G | +1992ApJS...78....1D | +1993AJ....106.1273Z |
| 1990MNRAS.243....1B | +1991ApJS...75..935d | 1992ApJS...78..365H- | 1993AJ....106.1743V- |
| 1990MNRAS.243..209M | +1991ApJS...75.1011G | 1992ApJS...79..157F | 1993AJ....106.2197T- |
| +1990MNRAS.243..390K | 1991ApJS...76...23W | +1992ApJS...79..255K- | +1993AN....314...97K |
| 1990MNRAS.243..504A- | 1991ApJS...76..471S- | +1992ApJS...79..331W | 1993AN....314..225S |
| 1990MNRAS.244..233R | 1991ApJS...76..813S | 1992ApJS...80....1S | 1993AN....314..317S |
| +1990MNRAS.244..408C | 1991ApJS...76.1043P | 1992ApJS...80..109B- | 1993Ap&SS.200..279H |
| 1990MNRAS.245..289I | 1991ApJS...76.1055D- | 1992ApJS...80..137J | 1993ApJ...404...81B |
| 1990MNRAS.246..110M | 1991ApJS...76.1067d- | 1992ApJS...80..211H | 1993ApJ...404..521A- |
| +1990MNRAS.246..169P | 1991ApJS...77..203J | 1992ApJS...80..257E | 1993ApJ...405..498W- |
| +1990MNRAS.246..256H | 1991ApJS...77..331H | 1992ApJS...80..479T | 1993ApJ...407..470Z- |
| +1990MNRAS.247..182B | +1991ApJS...77..363S | 1992ApJS...80..501O | 1993ApJ...409..110S |
| +1990PASJ...42..603S | 1991ChA&A..15...19X- | 1992ApJS...80..531F | 1993ApJ...411..501L- |
| 1990RMxAA..20....47C | +1991MNRAS.248..112W | 1992ApJS...81....5S | 1993ApJ...412..541D |
| 1991A&A...241...35B | 1991MNRAS.248..398W | 1992ApJS...81...35H | 1993ApJ...413..453W |
| +1991A&A...241..551W | *1991MNRAS.248..483S | 1992ApJS...81...83H | 1993ApJ...416...36P |
| 1991A&AS...87....1P | 1991MNRAS.248..804L- | 1992ApJS...81..413W | 1993ApJS...84..109L |
| 1991A&AS...87..309H | 1991MNRAS.249..164O- | 1992ApJS...82....1B | 1993ApJS...85....1W |
| 1991A&AS...87..389B | 1991MNRAS.249..606C- | 1992ApJS...82..417S | 1993ApJS...85..111C- |
| 1991A&AS...87..425R | +1991MNRAS.251...46H | 1992ApJS...83....1E | 1993ApJS...85..249H |
| 1991A&AS...88..559W- | *1991MNRAS.251..253B | +1992ApJS...83..299S | +1993ApJS...86....5K |
| 1991A&AS...89..389M- | +1991MNRAS.251..330W | 1992ApJS...83..650 | 1993ApJS...86..365P- |
| 1991A&AS...89..599H | 1991MNRAS.252..593Z- | 1992MNRAS.254...30G | 1993ApJS...87...63L |
| 1991A&AS...90...55v- | 1991MNRAS.253..222W | 1992MNRAS.254..655P | 1993ApJS...87..135O |
| 1991A&AS...90..121C | 1991MNRAS.253..584L- | 1992MNRAS.256..349Z- | +1993ApJS...87..451H |
| +1991A&AS...90..327t | 1991MNRAS.253..686C | +1992MNRAS.258....1L | 1993ApJS...88..383L |
| 1991A&AS...90..355C- | +1991Obs....111...75L | +1992MNRAS.259..233S | +1993ApJS...89....1R |
| 1991A&AS...90..375L | +1991PASAu...9..170O | 1992PASAu..10..140W | *1993ApJS...89...35G- |
| 1991A&AS...91....1P | 1991PAZh...17..787M- | +1992PASP..104..678S | 1993ApJS...89...57Y |
| 1991A&AS...91...61v- | +1991UCapT.11.....F | 1992PAZh...18..396A | *1993ApL...29....1H |
| 1991A&AS...91..285T | 1992A&A...260..355A | 1992RMxAA..24..147M- | *1993ESOSR.13......V |
| +1991A&AS...91..337S | 1992A&A...261...57B | +1993A&A...278....6X | *1993MNRAS.256..404L- |
| +1991A&AS...91..371P | 1992A&A...264..203G | +1993A&A...278..379B- | +1993MNRAS.260...77P |
| 1991A&AS...91..513R | 1992A&AS...92...63W | +1993A&AS...97..435Q | 1993MNRAS.261...39T |
| 1991AAA.53.002.023B | 1992A&AS...92..749T | 1993A&AS...98..165K- | 1993MNRAS.261..445F |
| +1991AAA.53.002.034d | 1992A&AS...93..173B | 1993A&AS...98..193V- | 1993MNRAS.262..475C |
| 1991AJ....101...18G | 1992A&AS...93..211F | 1993A&AS...98..229P- | +1993MNRAS.262.1057H |
| 1991AJ....101..127W | +1992A&AS...93..255G | 1993A&AS...98..275B | +1993MNRAS.263...25H |
| 1991AJ....101..148W | +1992A&AS...93..399W- | 1993A&AS...98..297B | 1993MNRAS.263...98B- |
| 1991AJ....101..362C | 1992A&AS...94..121T | 1993A&AS...99...71K- | 1993MNRAS.263..425M- |
| 1991AJ....101.1121H | 1992A&AS...94..299S | 1993A&AS...99..407C- | 1993MNRAS.263..707L- |
| 1991AJ....101.1183C | 1992A&AS...94..327T | 1993A&AS...99..545Z | 1993MNRAS.263..999T- |
| 1991AJ....101.1207G- | +1992A&AS...95....1G | 1993A&AS..100...33G- | 1993MNRAS.263.1023M |
| +1991AJ....101.1561P | 1992A&AS...95..129P- | +1993A&AS..100...47G | 1993MNRAS.264...71v- |
| 1991AJ....101.1983B | 1992A&AS...95..249L | 1993A&AS..100..263O | 1993MNRAS.264..298M- |
| 1991AJ....102..461C | 1992A&AS...95..489J- | 1993A&AS..100..453G- | 1993MNRAS.264..665V |
| 1991AJ....102.1258F- | 1992A&AS...96..389d | -1993A&AS..101..207v | 1993PASAu..10..310D |
| 1991AJ....102.1581B | 1992A&AS...96..435G- | -1993A&AS..101..447B | 1993PASJ...45.1530- |
| 1991AJ....102.1627W | 1992A&AS...96..441S | 1993A&AS..101..475Q | 1993PASP..105..387S |
| 1991AJ....102.1663C | 1992A&AS...96..583P | +1993A&AS..102...57B | +1993PNAOJ...3..169T |
| 1991AJ....102.1680W- | +1992AApTr...2..265K | -1993A&AS..102..251M | 1993PPMtO..12...65H |
| +1991AZh....68..681R | 1992AJ....103...11F | -1993A&AS..102..451M | 1993RMxAA..25...51M |
| +1991Afz....34....5S- | +1992AJ....103.1057V | +1993AAA.57.002.026G | +1994A&AS..103..245M |
| +1991Afz....34...13A | 1992AJ....103.1062L | 1993AJ....105...35C- | +1994A&AS..103..349S |
| *1991Afz....34..205S- | 1992AJ....103.1107S | 1993AJ....105...53B- | -1994A&AS..103..391M |
| +1991ApJ...368...54D | 1992AJ....103.1438R | 1993AJ....105..393K | +1994A&AS..103..573P |
| 1991ApJ...369...79L | 1992AJ....103.1501V | 1993AJ....105..427R | +1994A&AS..104....1F |
| 1991ApJ...375..150- | 1992AJ....103.1746N- | +1993AJ....105..788F | +1994A&AS..104..179G- |
| *1991ApJ...376...8F- | 1992AJ....104...1H- | 1993AJ....105..853I | -1994A&AS..104..259A |
| 1991ApJ...378...77G | 1992AJ....104..891F- | 1993AJ....105.1060P- | +1994A&AS..104..271G- |
| +1991ApJ...380...30M | 1992AJ....104.1706C | 1993AJ....105.1251W | -1994A&AS..104..529T |
| 1991ApJ...383..524Z- | +1992AN....313..189S | 1993AJ....105.1271G | +1994A&AS..105...67S- |
| +1991ApJS...75....1B | 1992AZh....69..225A | | +1994A&AS..105..211S |
| +1991ApJS...75..241D | +1992ApJ...384..404P | | +1994A&AS..105..385K |

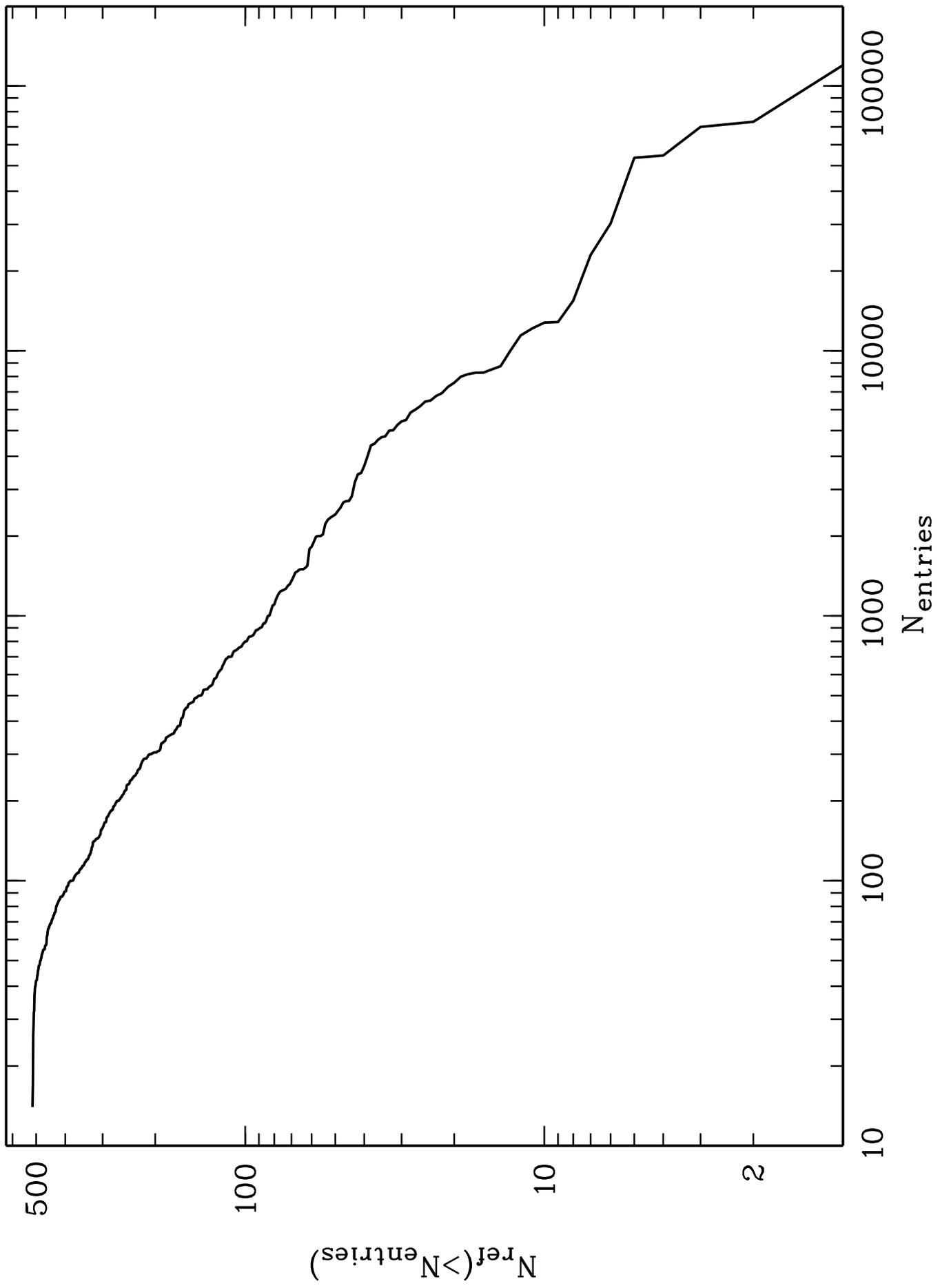

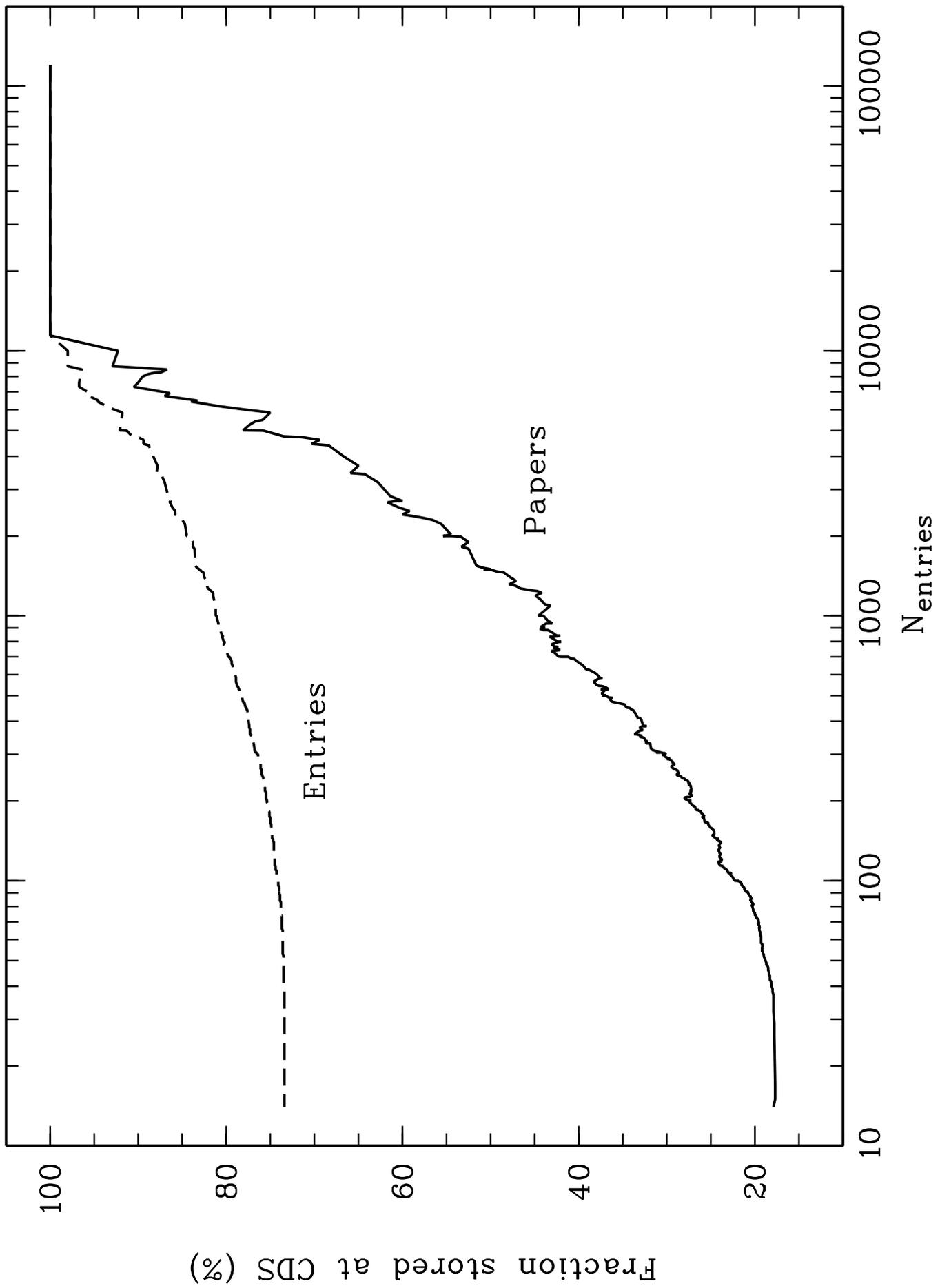